\documentclass{aa}
\usepackage{graphicx}

\begin{document}
   \title{Extinction calculations of multi-sphere polycrystalline graphitic clusters}

   \subtitle{A comparison with the 2175~{\AA} peak and between a rigorous 
            solution and discrete-dipole approximations}

   \author{Anja C.\,Andersen\inst{1}
          \and
          Juan A.\,Sotelo\inst{2,}\inst{4}
          \and
          Vitaly N.\,Pustovit\inst{3,}\inst{4}
          \and
          Gunnar A.\,Niklasson\inst{4}
          }

   \offprints{A.C.\,Andersen}

   \institute{Department of Astronomy \& Space Physics, 
     Uppsala University, P.O.Box 515, SE-751 20 Uppsala, Sweden\\
   \email{anja@astro.uu.se}
   \and 
     Dpto. de Fisica, Informatica y Matematicas, Universidad Peruana Cayetano Heredia,
     Aptdo. 4314, Lima, Peru \\
   \email{juan.sotelo@angstrom.uu.se} 
   \and 
     Institute of Surface Chemistry, NAS of Ukraine, 17 Gen.\ Naumova str., 
      Kiev 03164, Ukraine \\
   \email{pustovit.vitaly@angstrom.uu.se} 
   \and
     Department of Materials Science,  
     Uppsala University, P.O.Box 534, SE-751 21 Uppsala, Sweden\\
   \email{gunnar.niklasson@angstrom.uu.se} 
   }
   \date{Received 15/10-2001; accepted 8/1-2002}

   \abstract{Certain dust particles in space are expected to appear as clusters
of individual grains. The morphology of these clusters could be fractal or compact. 
In this paper we study the extinction by compact and 
fractal polycrystalline graphitic clusters consisting of touching identical spheres, based on the dielectric function of graphite from Draine \& Lee (\cite{draine+lee}).
We compare three general methods for computing the extinction of the clusters 
in the wavelength range $0.1 - 100~\mu$m,
namely, a rigorous solution (G\'erardy \& Ausloos \cite{GA}) and two different discrete-dipole
approximation methods -- MarCODES (Markel \cite{markel}) and DDSCAT 
(Draine \& Flatau \cite{draine+flatau}). 
We consider clusters of $N=$ 4, 7, 8, 27, 32, 49, 108 and 343
particles of radii either 10~nm or 50~nm, arranged in three
different geometries: open fractal (dimension D $= 1.77$), simple cubic and
face-centred cubic. 
The rigorous solution shows that the extinction of the fractal clusters, 
with $N \le 50$ and particle radii 10~nm, displays a peak within 2\% 
of the location of the observed interstellar extinction peak at
$\sim 4.6 \mu$m$^{-1}$; the smaller the cluster, the closer its peak gets to this
value. By contrast, the peak in the extinction of the more compact
clusters lie more than 4\% from $4.6 \mu$m$^{-1}$.
At short wavelengths ($0.1 - 0.5 \mu$m), all the methods show
that fractal clusters have markedly different extinction from those of
non-fractal clusters. At wavelengths $> 5 \mu$m, 
the rigorous solution indicates that the extinction from fractal and
compact clusters are of the same order of magnitude.  
It was only possible to compute fully converged results of the rigorous
solution for the smaller clusters, due to computational limitations, 
however, we find that both discrete-dipole approximation methods 
overestimate the computed extinction of the smaller fractal clusters. 
    }
   \authorrunning{Andersen et al.} 
   \titlerunning{Extinction calculations of multi-sphere polycrystalline graphitic clusters}
   \maketitle
%
%________________________________________________________________

\section{Introduction}

The shape of interstellar and circumstellar grains is still an outstanding 
issue. For many years, the complexity of the electromagnetic scattering 
problem to solve has limited the shapes studied to spheres, infinite 
cylinders and spheroids. However, the shape of many interstellar
grains are expected to be non-spherical and maybe even highly irregular.
One way to deal with irregular particles and so with clusters of
dust grains is to assume that they consist of touching spheres.
With such an assumption it is possible to construct many distinct different
morphologies which can then be compared with observations. 

The problem of evaluating the extinction efficiency ($Q_{\rm ext}$) is that of
solving Maxwell's equations with appropriate boundary conditions at the 
cluster surface. A solution was formulated by Lorenz (\cite{lorenz}) and Mie (\cite{mie})
for a homogeneous single sphere and the complete formalism is therefore often 
referred to as Lorenz-Mie theory. A complementary solution based on 
expansion of scalar potentials was given by Debye (\cite{debye}). A detailed
description of this exact electromagnetic solution can be found in
Bohren \& Huffman (\cite{bohren+huffman}).  For a comprehensive review on
the optics of cosmic dust see Voshchinnikov (\cite{nicolai02}).

An updated overview of exact theories and numerical techniques for computing
the scattered electromagnetic field by clusters of particles is given
in (Mishchenko et al. \cite{Mish1}; \cite{Mish2}; Fuller \& Mackowski 
\cite{Fuller}; Ciric \& Coorey \cite{Ciric}; Draine \cite{draine00};
Voshchinnikov \cite{nicolai02}  and references therein). All
of these methods are based on solving Maxwell's equations.
For clusters of spheres embedded in non-absorbing
media one such method is based on the G\'erardy and Ausloos theory 
(G\'erardy \& Ausloos \cite{GA4}; \cite{GA}; \cite{GA2}; \cite{GA3}) 
which recently has been extended to also treat clusters embedded in absorbing media 
(Lebedev \& Stenzel \cite{lebedev1}; Lebedev et al. \cite{lebedev2}).
Another method that is often used in practice is the discrete dipole approximation (DDA);
it has been used in a wide range of scattering problems concerning clusters
of particles including the extinction of
interstellar dust grains (e.g.\ Vaidya et al. \cite{vaidya}; Bazell \& Dwek \cite{bazell}).

The Galactic interstellar extinction curve displays a ``2175~{\AA} peak'',
which has presented an astrophysical puzzle since its discovery by Stecher (\cite{stecher}).
Fitzpatrick \& Massa (\cite{fitzpatrick+massa86}) studied the interstellar
extinction in the direction of 45 reddened stars, and found that it displays
a peak whose central wavelength $\lambda_{0}$ is remarkably constant
($\lambda_{0} = 2174.4 \pm 17$ {\AA}), even though its full width at half
maximum (FWHM) varies considerably from $~360$ to $~600$ {\AA}; they also
found no apparent correlation between the small variation in $\lambda_{0}$
and the large variation in the FWHM. 

Graphite has long been considered a very promising though controversial
candidate for explaining the 2175~{\AA} peak 
(e.g.\ Stecher \& Donn \cite{stecher+donn}; 
Fitzpatrick \& Massa \cite{fitzpatrick+massa}; Mathis \cite{mathis}; Voshchinnikov \cite{nicolai90};
Sorell \cite{sorell}; Draine \& Malhotra \cite{draine+malhotra}; 
Rouleau et al. \cite{rouleau}; Will \& Aannestad \cite{will+aannestad}). 
It is considered a promising candidate because small, 0.015 $\mu$m,
graphitic spheres display an extinction peak at about the right wavelength
with a FWHM in accordance with the observations
(Gilra \cite{gilra}), and because its abundance do not seem to contradict 
the cosmic abundance constraints (Snow \& Witt \cite{snow+witt}). 
It is considered a controversial candidate, however, because 
increasing the size of a small graphitic particle simultaneously increases
the central wavelength and FWHM of its extinction peak. 
The peak will shift to longer wavelengths when the particle size is increased,
when the particle shape is oblate spheroidal and   
when the particles are coated with a dielectric substance such as ice;
it will shift to shorter wavelengths when the particle shape is prolate spheroidal
(Gilra \cite{gilra}; Hecht \cite{hecht}; Draine \& Malhotra \cite{draine+malhotra}).
The observed lack of correlation
between the position and FWHM of the peak therefore presents a challenge to
the hypothesis that graphite particles originate this peak.
In an extensive investigation, Draine \& Malhotra (\cite{draine+malhotra}) 
conclude that if graphite particles are the carriers of the 2175~{\AA} peak, a 
variation in their optical properties ought to be present. This may be the result 
of varying amounts of impurities, variations in crystallinity, or changes in
its electronic structure due to surface effects. 
Clustering effects have also been considered;
Rouleau et al. (\cite{rouleau}) have shown that compact clusters of graphitic
spheres satisfy the criterion that the position of the peak
remain stable while the width varies. 

To investigate the clustering effect further we here compute and analyse the extinction 
of different polycrystalline graphitic clusters. We have chosen clusters 
ranging from small to large, and that are either sparse or compact.  
In this way we intend to evaluate how the extinction is
influenced by the structure. 
We focus on clusters consisting of 4, 7, 8, 27, 32, 49,
108 and 343 touching polycrystalline graphitic spheres. 
The extinction of the clusters is
calculated by the use of a rigorous (GA) method (G\'erardy \& Ausloos \cite{GA}) 
as well as by two DDA methods -- MarCODES 
(Markel \cite{markel}) and DDSCAT (Draine \& Flatau \cite{draine+flatau}) -- 
 to test how well these approximations
perform when applied to clusters with different morphology.  
Recently Xu \& Gustafson (\cite{xu+gustafson}) compared
light-scattering calculations by a rigorous solution similar to the
GA method, with DDSCAT for two identical polystyrene spheres in contact.
They found that DDSCAT worked reasonably well on small volume structures while
its validity is challenged on large structures. We find a tendency for both
the DDA codes to overestimate significantly the extinction for small 
(N $< 10$) fractal clusters and to a much larger extent than for the comparable
compact clusters.

By studying clusters consisting of polycrystalline graphitic particles we deal with
the anisotropy of graphite in a different way than usual (e.g. Draine 
\cite{draine88})  and 
it turns out that this has an influence on the comparison of
the calculated extinction with the observed interstellar extinction.

This paper is structured as follows. Section \ref{graphite_sect}  deals with the
anisotropy of graphite. Section \ref{structure_sect}
introduces the fractal and compact clusters studied in this work.
Section \ref{computationalmethods_sect} describes the computational methods
under scrutiny; the exact GA solution (G\'erardy \& Ausloos \cite{GA}) and the
two different DDA methods, developed by Draine \& Flatau (DDSCAT, \cite{draine+flatau}) 
and Markel (MarCoDES, \cite{markel}). Section \ref{discussion_sect} 
presents and dicusses our results for clusters from one up to 343 particles, 
in the wavelength range $0.03 - 100~\mu$m.
Section \ref{fuv_sect} discusses the implications of our results in 
the interpretation of the interstellar extinction curve.
Section \ref{conclusion_sect} presents the conclusions.

%__________________________________________________________________

\section{Graphite} \label{graphite_sect}

Theoretical computation of absorption and scattering by graphitic particles
is difficult because graphite is a semi-metal with high anisotropy.
Graphite can be characterised by two different dielectric functions, 
$\epsilon_{\perp}$ and $\epsilon_{\parallel}$, corresponding to the
electric-field vector $\mathbf{E}$ being perpendicular ($\epsilon_{\perp}$) and
parallel ($\epsilon_{\parallel}$) to the symmetry axis of the crystal ($c$-axis), which 
is perpendicular to the basal plane. 
It is a lot
easier to experimentally determine $\epsilon_{\perp}$ than  $\epsilon_{\parallel}$, 
because graphite cleaves readily along the basal plane and hence reflectivity
measurements can be made with normally incident light, whereas
it is very difficult to prepare suitable optical surfaces 
parallel to the $c$-axis. 

In this work we deal with the anisotropy of graphite by assuming that in
all our clusters, each individual particle is polycrystalline having a
dielectric function $\epsilon_{\rm ave}$ given by the arithmetic average of
$\epsilon_{\parallel}$ and $\epsilon_{\perp}$, namely 
$\epsilon_{\rm ave} = \frac{1}{3} \epsilon_{\parallel} + \frac{2}{3} \epsilon_{\perp}$.
For a polycrystal, this arithmetic average is in fact an attainable upper bound for 
its dielectric function (Avellaneda et al. \cite{avellaneda}).
We use the dielectric functions $\epsilon_{\parallel}$ and
$\epsilon_{\perp}$ derived by Draine \& Lee (\cite{draine+lee}) covering 
the region from the far-IR to the
far-UV.  The data agree well in the 
0.03$-$62~$\mu$m region with those published by Borghesi \& Guizzetti
(\cite{borghesi+guizzetti}). 

In constrast, the usual ``1/3$-$2/3'' approximation, treats individual particles as
mono-crystalline - 1/3 of the cluster particles are assumed to have 
dielectric function $\epsilon_{\parallel}$ and the remaining 2/3 
to have dielectric function $\epsilon_{\perp}$. This approximation has been shown
by Draine (\cite{draine88}) and Draine \& Malhotra (\cite{draine+malhotra}) 
to have a surprisingly good accuracy for graphite grains with radii $\leq 200$~{\AA}.  
However, considering
grain formation and grain growth in stellar environments (Sedlmayr \cite{sedlmayr}),
mono-crystalline particles do not seem to be as valid an assumption as 
polycrystalline ones. Also, investigations of presolar nano-diamond grains from 
meteorites -  direct specimens of surviving physical material formed in past 
stellar environments which can be quantitatively analysed in the
laboratory - show a clear tendency towards being polycrystalline (Daulton 
et al. \cite{daulton}; Phelps \cite{phelps}).

%
%                                                One column figure
%----------------------------------------------------------- 
   \begin{figure}
   \centering
     \includegraphics[angle=90,width=9cm,clip]{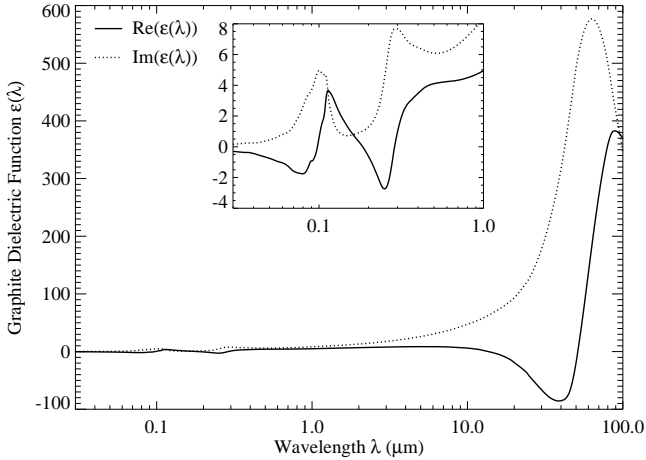}
     \caption{The average 
      dielectric function ($\epsilon_{ave} = \frac{2}{3} \epsilon_{\perp} + 
     \frac{1}{3} \epsilon_{\parallel}$; see text) of graphite based on the dielectric functions
    derived by Draine \& Lee (\cite{draine+lee}).}
         \label{diel_grap}
   \end{figure}
%
%______________________________________________________________

A particle in vacuum will show a Lorenz-Mie absorption peak whenever
the real part of its dielectric function, $\Re (\epsilon)$, satisfies 
$\Re (\epsilon) = -2$. Looking at the average dielectric function of graphite 
shown in Fig.\,\ref{diel_grap}, absorption peaks should occur at about $0.220~\mu$m
and $14.6~\mu$m. The latter should be much more damped due to the higher value 
of the imaginary part of $\epsilon$.

\section{Structure of the clusters} \label{structure_sect}

We consider three-dimensional clusters of identical touching spherical particles
arranged in three different geometries: fractal,
simple cubic, and face-centred cubic. 
The structures do not have shapes expected to be
found in space, but will provide us with boundary conditions
for the problem of calculating the extinction from clusters of grains
of different morphology. Table\,\ref{clusters} lists all the clusters used in this work.

\begin{table}
\caption{The clusters presented in this paper have three
different geometries: fractal (frac; D$=1.77$), face-center cubic 
(fcc) and simple cubic (sc).}
\begin{center}
\begin{tabular}{|c|c|c|} \hline
Cluster & \# particles & Designation in \\
structure  & in cluster & this paper  \\ \hline
frac & 7 & frac7  \\
frac & 49 & frac49  \\
frac & 343 & frac343 \\
fcc   & 4 & fcc4  \\
fcc & 32 & fcc32  \\
fcc & 49 & fcc49  \\
fcc & 108 & fcc108  \\
sc & 8 & sc8  \\
sc & 27 & sc27  \\ \hline
\end{tabular}
\end{center}
\label{clusters}
\end{table}

\begin{figure}
   \begin{center}
    \includegraphics[angle=90,bb=30 120 550 700,width=9cm,clip]{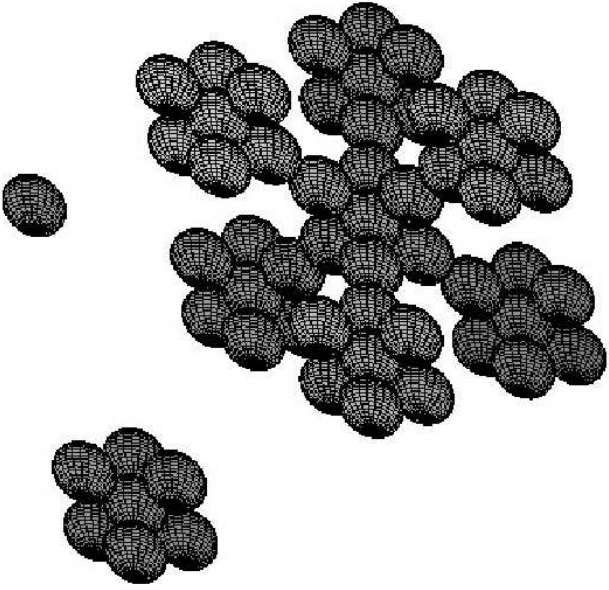}
     \caption{Stages of fractal construction.
      The seed particle, the generator (quasi-fractal clusters 
         consisting of 7 particles) and the 3-D structure of the
       49 particle fractal cluster (D$=1.77$), see text for more details.}
         \label{frac}
   \end{center}
   \end{figure}

The fractal clusters, frac7, frac49, and frac343, are obtained from the first 
three stages of the recursive construction
of the snowflake fractal. This construction can be summarised as follows: 
a seed particle is put at the 
origin of the coordinate system. In the first step a generator is 
built by symmetrically gluing to the seed particle six copies of itself along 
the x, y, and z axes. Next, each particle in the first configuration, 
the generator, is substituted by the whole generator itself. In the next 
steps the same rule is applied: each particle is replaced by the generator.
Fig.\,\ref{frac} shows this procedure up to the second stage.
For a fractal cluster its dimension $D$ is defined by 
\begin{equation}
M(r) = M_{0} \left[ \frac{r}{r_{0}} \right] ^{D},
\end{equation}
where $M(r)$ is the mass of material contained within a sphere of radius $r$.
For a solid grain of constant density $D=3$, whereas for a fractal grain $D<3$
(Mandelbrot \cite{mandelbrot}); in particular, for the snowflake fractal 
$D = {\rm ln 7} / {\rm ln 3} = 1.77$ (Vicsek \cite{vicsek}). 

Although such a deterministic structure (Fig.\,\ref{frac}) 
is not expected to occur in nature, its 
fractal dimension is close to that of more 
realistic random cluster-cluster aggregation models. In particular, small particles in space may move in straight, 
ballistic trajectories and form larger aggregates upon collisions. Numerical 
simulation of this process yields a value of around 1.9 for the fractal 
dimension of the resulting 
aggregates (Meakin \cite{meakin}; Botet and Jullien \cite{botet};
Meakin \& Jullien \cite{meakin+jullien}); this value has also been confirmed
by experimental results (Wurm \& Blum \cite{wurm}). Moreover,
optical properties of fractal clusters with $D<2$ are predicted to be significantly 
different from those with $D>2$ (Berry \& Percival \cite{berry}); 
hence we consider important to use a fractal with a realistic 
dimension in our computations. 
The open fractal structure presents a challenge for modelling due to the high 
porosity and large surface area.

As a contrast to the fractal structure, compact crystalline structures are 
studied, namely, face-centred cubic and simple cubic, 
see e.g. Kittel (\cite{kittel}) for a discussion on crystal 
structures. All the clusters listed in Table\,\ref{clusters} are symmetric 
except the fcc49\footnote{For such 
an fcc cluster 48 particles would form a $3 \times 2 \times 2$ unit-cell,
while 49 particles can only be fitted into a $3 \times 2 \times 3$ unit-cell.}. 
The non-fractal clusters are a challenge to model since they are far from being
spherical.

%
%                                                One column figure
%----------------------------------------------------------- 
%
%______________________________________________________________

\section{The computational methods} \label{computationalmethods_sect}

\subsection{G\'erardy and Ausloos theory} \label{ga}

A rigorous and complete solution to the multi-sphere light scattering 
problem has been given by G\'erardy \& Ausloos (GA) (\cite{GA4};
\cite{GA}; \cite{GA2}; \cite{GA3}) as
an extension of the Mie-Ruppin theory (Mie \cite{mie}; Ruppin \cite{ruppin}). 
It is based on the
exact solution of Maxwell's equations for arbitrary cluster geometries,
polarisation  and incidence direction of the light. This is done by expanding
the various
fields involved in terms of the vector spherical harmonics (VSH).
The usual boundary conditions are extended to take into account the possible
existence of longitudinal plasmons\footnote{A collective excitation for
quantized plasma oscillations, in which the free electrons in a metal are
treated as a plasma.} in the spheres.
High-order multipolar electric and magnetic interaction
effects are included. We summarise their work as used throughout the paper.

We consider a cluster of $N$ homogeneous spheres of radius $R$ and
dielectric function $\epsilon(\omega)$, embedded in a matrix of dielectric
constant $\epsilon_M$ and submitted to a plane polarised time harmonic
electromagnetic field. The total scattered field from the cluster is
represented as a superposition of individual fields scattered from each
sphere. The electromagnetic field impinging on each sphere
consists of the external incident wave and the waves scattered by the other
spheres. For any sphere, the incident, internal and scattered fields are
expressed in VSH centred at the sphere origin.
The boundary conditions on its surface are
solved by transforming all relevant field expansions into the sphere
coordinate system, yielding the following system of linear coupled equations
for the expansion coefficients $c_{\mu}$ and $d_{\mu}$ of the scattered
field (G\'erardy \& Ausloos \cite{GA}):
\begin{eqnarray}
c_{\mu} &= \Gamma_q\,\{\, {a_0}_{\mu} + \sum_{\nu}^{'}
( J_{\mu\nu}\ c_{\nu} + C_{\mu\nu}\ d_{\nu} )\, \}, \label{juan1} \\
d_{\mu} &= \Delta_q \{\, {b_0}_{\mu} + \sum_{\nu}^{'}
( C_{\mu\nu}\ c_{\nu} + J_{\mu\nu}\ d_{\nu} )\, \}, \label{juan2}
\end{eqnarray}
where $\mu=(q,p,i)$ and $\nu=(n,m,j)$. The indices $q$ and $n$ denote the
polar order, $p=-q,\dots, q,\ m=-n,\dots, n$, and the indices $i, j$ number
the particles in the cluster. Moreover, ${a_0}_{\mu}$ and ${b_0}_{\mu}$ are
the coefficients of the expansion of the external incident wave in VSH in
the coordinate frame of the $i$~th sphere (G\'erardy \& Ausloos \cite{GA}). The multi-polar electric
and magnetic susceptibilities of a sphere (G\'erardy \& Ausloos \cite{GA}) are denoted by
$\Delta_q$ and $\Gamma_q$ respectively. The coefficients $J_{\mu\nu}$ and
$C_{\mu\nu}$
describe the transformation of the VSH from a frame centred on particle $j$
to another centred on particle $i$. Analytical expressions for these
coefficients have been given by G\'erardy and Ausloos (\cite{GA}). The primes
in the sums in Eqs.\,(\ref{juan1}) and (\ref{juan2}) indicate that terms with
$j=i$ are omitted. The solution to this system of equations can be written
in matrix form as

\begin{equation}
\left[ 
{\begin{array}{c}
\mathbf{c} \\ \mathbf{d} 
\
\end{array}}
\right]
=
\left[ 
{\begin{array}{cc}
\mathbf{T}_{N11} & \mathbf{T}_{N12} \\
\mathbf{T}_{N21} & \mathbf{T}_{N22}
\end{array}}
\right]
\left[ 
{\begin{array}{c}
\mathbf{a_0} \\ \mathbf{b_0}
\end{array}}
\right] ,
\label{juan3}
\end{equation} 
where $\mathbf{T}_{N}$ is the T-matrix (Waterman \cite{waterman};
Tsang et al. \cite{tsang}; Mishchenko et al. \cite{Mish2}) of the cluster with components
\begin{eqnarray} 
\mathbf{T}_{N11} &=& (\,(\mathbf{\Gamma}^{-1}-\mathbf{J}) -
\mathbf{C} (\mathbf{\Delta}^{-1}-\mathbf{J})^{-1} \mathbf{C}\,)^{-1} \\ \nonumber
\mathbf{T}_{N12} &=& \mathbf{T}_{N11}
\mathbf{C}\,(\mathbf{\Delta}^{-1}-\mathbf{J})^{-1} \\ \nonumber
\mathbf{T}_{N21} &=& (\mathbf{\Delta}^{-1}-\mathbf{J})^{-1} \mathbf{C}\,
\mathbf{T}_{N11} \\ \nonumber
\mathbf{T}_{N22} &=& (\mathbf{\Delta}^{-1}-\mathbf{J})^{-1} (\mathbf{C}\,
\mathbf{T}_{N12} + \mathbf{I}),
\end{eqnarray}
and $\mathbf{I}$ is the identity matrix.  When $N=1$, Eq.\,(\ref{juan3}) is
just the Mie-Ruppin result (Mie \cite{mie}; Ruppin \cite{ruppin})
\begin{equation}
\left[ 
{\begin{array}{c}
\mathbf{c} \\ \mathbf{d}
\end{array}}
\right]
=
\left[ 
{\begin{array}{cc}
\mathbf{\Gamma} & \mathbf{0}\\
\mathbf{0} & \mathbf{\Delta}
\end{array}}
\right]
\left[ 
{\begin{array}{c}
\mathbf{a_0} \\ \mathbf{b_0}
\end{array}}
\right] .
\label{juan4}
\end{equation}

The extinction efficiency, $Q_{ext}$, i.e. the extinction cross-section in
units of the total geometrical cross-section can then be obtained
from G\'erardy \& Ausloos (\cite{GA})
\begin{equation}
Q_{ext} = -\frac{1}{\pi Nk^2R^2}\,\Re(\mathbf{a}_0^{*T}\mathbf{c} +
\mathbf{b}_0^{*T}\mathbf{d}),
\label{juan5}
\end{equation}
where $k=2\pi /\lambda$ and $\lambda$ is the wavelength in the matrix.
The extinction cross section per unit volume is then computed as
\begin{equation}
\frac{C}{V} = \frac{3}{4R} Q_{ext}.
\label{juan6}
\end{equation}

Limiting $q$ and $n$ in Eqs.\,(\ref{juan1}) and (\ref{juan2}) 
to integers less than or equal to
$L$, we obtain a system of $2NL(L+2)$ equations whose solution, 
Eq.\,(\ref{juan3}), is the
$2^L$-polar approximation to the electromagnetic response of the cluster;
$\mathbf{a}_0,\,\mathbf{b}_0,
\,\mathbf{c}\,\mbox{ and }\mathbf{d}$ are complex vectors of dimension
$NL(L+2)$, while the $\mathbf{T}_{Nij},\,i,j=1,2$, are complex
square matrices of dimension $NL(L+2)$.
In this case Eqs.\,(\ref{juan5}) and (\ref{juan6}) determine the $2^L$-polar
approximation to the cluster extinction efficiency and the cluster extinction 
per unit volume respectively. 

The extension of this theory to treat the case of
absorbing embedding media is given in (Lebedev \& Stenzel \cite{lebedev1};
Lebedev et al. \cite{lebedev2}).

\subsection{The DDA method}

The discrete dipole approximation (DDA) - also known as 
the coupled dipole approximation - method is one of several discretisation 
methods
(e.g. Draine \cite{draine88}; Hage \& Greenberg \cite{hage+greenberg}) 
for solving scattering 
problems in the presence of a target with arbitrary geometry.
The discretisation of the integral form of Maxwell's equations 
is usually done by the method of moments (Harrington \cite{harrington}).
Purcell \& Pennypacker (\cite{purcell}) were the first to 
apply this method to astrophysical problems; 
since then, 
the DDA method has been improved greatly by Draine (\cite{draine88}), 
Goodman et al. (\cite{goodman}), Draine \& Goodman (\cite{draine+goodman}), 
Draine \& Flatau (\cite{draine+flatau}), 
Markel (\cite{markel}), and Draine (\cite{draine00}). The
 DDA method has gained popularity among 
scientists due to its clarity in physical principle and the  
FORTRAN implementation which have been made publicly available by Draine \& 
Flatau (DDSCAT
package, Draine \& Flatau \cite{draine+flatau}) and by Markel (MarCoDES,
 Markel \cite{markel}). 

When considering the problem of scattering and absorption of linearly polarised
light
\begin{equation}
E_{0} = e_{0} \exp \left( ikr \right) \label{vitaly1}
\end{equation}
by an isotropic graphitic grain, then 
within the concept of the DDA method, the grain
is replaced by a set of discrete elements of volume $V_i$ with relative
dielectric constant $\epsilon_i$ and dipole moments $d_i=d\left(
r_i\right)$, $i=1,...,N,$ whose coordinates are specified by vectors $r_i$.
The equations for the dipole moments can be written
using simple considerations based on the concept of the exciting field,
which is equal to the sum of the incident wave and the fields of the rest of the dipoles in a given point
\begin{equation}
\mathbf{d}_{i} \left( r_{i} \right) = \alpha_{i} \left[ E_{0} \left( r_{i} 
     \right) + k^{3}
\sum_{j\neq i} \mathbf{G}_{ij} \mathbf{d}_{j} \left( r_{j} \right) \right]  \label{vitaly2}
\end{equation}
where the dipole scattering tensor $\mathbf{G}_{ij}$ has the following form in dyad
notations:
\begin{eqnarray}
\mathbf{G}_{ij} &=&\left[ A\left( kr\right) \delta _{ij}+B\left( kr\right) \frac{%
r_ir_j}{r^2}\right]  \label{vitaly3} \\
A\left( x\right) &=&\left[ x^{-1}+ix^{-2}-x^{-3}\right] \exp \left( ix\right)
\nonumber \\
B\left( x\right) &=&\left[ -x^{-1}-3ix^{-2}+3x^{-3}\right] \exp \left(
ix\right)  \nonumber \\
\mathbf{r} & \equiv & \mathbf{r}_{ij} = \mathbf{r}_{i} - \mathbf{r}_{j}. \nonumber
\end{eqnarray}
The solution of the set of Eqs.\,(\ref{vitaly2}) and (\ref{vitaly3}) 
yields all the basic optical
characteristics of a particle such as the integrated extinction $\left(
Q_{\rm ext}\right)$ and absorption $\left( Q_{\rm abs}\right) $:

\begin{eqnarray}
Q_{\rm ext} &=& 4 \pi k \Im \left[ \sum_{i} \left( e_{0} d_{i} \right) \exp (-ik_{0}r_{i}) \right]
\label{vitaly4} \\
Q_{\rm abs} &=& 4 \pi k\sum_{i} \eta _{i} \left| d_{i} \right| ^2,\hspace*{0.4 cm} 
\eta_{i} = \frac{4 \pi \Im \left( \epsilon _{i} \right) }{V_{i} \left| 
    \epsilon_{i-1} \right| ^2},  \nonumber 
\end{eqnarray}
and scattering $\left( Q_{\rm sca}=Q_{\rm ext} - Q_{\rm abs} \right)$ efficiencies. Here $\Im$ means the
imaginary part of the expression in the argument.

Once the location and
polarisability of the points are specified, calculations of the 
scattering and absorption of light by the 
array of polarisable points can be carried out to in principle whatever 
accuracy is required. The limiting factor is the 
capacity of computing resources.

\subsubsection{DDSCAT} \label{ddscat}

In this work we use the DDSCAT code version 5a10 (Draine \& Flatau \cite
{draine+flatau}; Draine \& Flatau \cite{draine+flatau00}). 
This version contains a new shape option where a target can be defined as the 
union of the volumes of an arbitrary number of spheres. 
In DDSCAT the considered grain/cluster is replaced by a cubic array
of point dipoles. The cubic array has numerical advantages because the
conjugate gradient method can be efficiently applied to solve the matrix
equation describing the dipole interactions (Goodman et al. \cite{goodman}).
When knowing the dipole strength of each cell in the particle it is possible to
 compute the
optical properties of arbitrary dust configurations.

There are three criteria for validity of DDSCAT:
\begin{enumerate}
\item  The wave phase shift $\rho =|m|kd$ ($m = \sqrt{\epsilon}$ being the
complex refractive index of the target material) over the distance $d$
between neighbouring dipoles should be less than 1 for calculations of total
cross sections and less than 0.5 for phase function calculations.
\item  $d$ must be small enough to describe the object shape satisfactorily.
\item The refractive index $m$ must fulfill $|m| < 2$.
\end{enumerate}

A comparison study by Draine \& Flatau (\cite{draine+flatau}) shows that
scattering and absorption cross sections can be calculated with DDSCAT to
accuracies of a few percent, provided that the
criteria elaborated above are satisfied.

Xing \& Hanner (\cite{xing+hanner}) state that the $Q$ efficiency factor is
generally much less sensitive to those criteria, and that it is usually
sufficient to let the value of $\rho$ be around 1 or possible bigger. 
Xu \& Gustafson (\cite{xu+gustafson}) show through a comparison between the
exact solution for two spheres in contact and DDSCAT that the criterion set
up by Draine \& Flatau (\cite{draine+flatau}) is not sufficient to ensure
high accuracy if the particles are large and strongly interacting (high
refractive index). They, therefore, recommend to use a validity criteria of 
$\rho < 0.3$ when calculating total cross sections.

The typical number of dipoles needed to obtain a reliable computational
result using the DDSCAT code can be determined by calculating the minimum
number of dipoles needed per particle. When a particle is represented by a
3-dimensional array of $N$ dipoles, its volume is $Nd^{3}$, which
must be equal to $4\pi R^{3} / 3$,
\begin{equation}
N = \frac{4 \pi}{3} \left( \frac{R}{d} \right)^{3}  
  = \frac{4 \pi}{3} \left( \frac{2 \pi R |m|}{\rho \lambda} \right)^{3} 
  \approx 1039 \left( \frac{R |m|}{\rho \lambda} \right)^{3}
\label{hanner}
\end{equation}
since $d$ is related to the wave phase shift $\rho$ by $d = \rho / (|m|k)$
(Draine \& Flatau \cite{draine+flatau00};
Xing \& Hanner \cite{xing+hanner}). 
The number of dipoles needed to satisfactorily describe the shape of the
object considered might be larger than indicated by Eq.\,(\ref{hanner}) if the
shape is ``challenging''. It is important to remember that Eq.\,(\ref{hanner}) 
only fulfills the first of the three criteria set by Draine \& Flatau
(\cite{draine+flatau00}).

For materials with large refractive index ($|m| > 2$), Draine \& Goodman
(\cite{draine+goodman}) show that especially the absorption is overestimated 
by DDA.  The limitation in DDSCAT is set by the use of
the lattice dispersion relation (LDR) 
for electromagnetic waves propagating on an 
infinite cubic lattice of point dipoles of polarisability $\alpha_{i}$ and
spacing $d$. According to Draine \& Goodman (\cite{draine+goodman}), the LDR 
prescription for $\alpha_{i}$ gives fair accuracy for scattering but poorer
results for absorption. When $|m - 1|$ is large the continuum material 
is effective at screening the external field: 
in the limit $|m - 1| \rightarrow \infty$
the internal field generated by the polarisation would exactly cancel the
incident field, so that the continuum material in the interior of the target
would be subjected to zero field. In the case of a discrete dipole array,
the dipoles in the interior will also be effectively shielded, while 
the dipoles located on the target surface are not fully shielded and, 
as a result, absorb energy from the external field at an excessive rate. 
This error can be
reduced to any desired level by increasing the number $N$ of dipoles,
thereby minimising the fraction $~ N^{-1/3}$ of the dipoles which are at 
surface sites, but very large values of $N$ are required when $|m|$ is
large (Draine \& Goodman \cite{draine+goodman}).

Graphite is characterised by having a high refractive index.
This means that the criterion $|m|<2$ is only
fulfilled shortward of 0.072 $\mu $m ($m=n+ik=0.58+i1.42$) and
relaxing the criterion a little (to account for the variability 
of $m$ that occurs)
gives an upper limit of 0.216~$\mu$m ($m=0.66+i1.35$).

\subsubsection{MarCoDES} \label{marcodes}

Another efficient code based on DDA is the Markel Coupled
Dipole Equation Solver (MarCoDES, Markel \cite{markel}). This code is
designed to solve the coupled Eq.\,(\ref{vitaly2}) for an arbitrary cluster
of $point$ dipoles using either conjugate gradient method (iterative method)
or the LU expansion (direct method). The program is in principle applicable
to clusters of small particles with arbitrary geometry, but is most
computationally efficient for sparse clusters (i.e. when the volume fraction
is very low) with significant number of particles $(\approx 10^3-10^4)$. Contrary
to DDSCAT the program does not use the Fast Fourier Transformation (FFT)
because this might significantly decrease the computational performance for
clusters with a low volume filling factor. When the volume filling fraction
is close to unity, algorithms utilising FFT will be much faster. In the
framework of the MarCoDES a spherical grain is considered to
be equivalent to the single elementary dipole, this corresponds to $N=1$ in
the DDSCAT case. The dimensionality of the coordinates of particles in
MarCoDES require a special consideration. By replacing real particles by
point dipoles located at their centres we significantly underestimate the
strength of their interaction. In order to correct the interaction strength,
the author of MarCoDES introduces geometrical intersection of particles.
All coordinates are defined  in terms of the distance between neighbouring 
dipoles $d$, which is given by $d = \left( 4\pi /3\right)^{1/3} R$. 
So, for example, if two particles have radii 10~nm, then the
distance between the dipoles is 16.12~nm. This suggested
phenomenological procedure allows MarCoDES to be 
more accurate than the 
usual single dipole approximation since the intersection
produces some
analogy of including higher multipole interactions between particles.
Application of MarCoDES to fractal sparse clusters allows, in principle, to
introduce renormalisation on the sphere radii and number of particles
depending on the fractal dimension of the cluster 
(Markel et al. \cite{markel00}), which will
produce even further corrections to the strength. In this paper we use the
simple particle intersection model given in the MarCoDES. 
The fact that the program only uses a single dipole for each particle in the
cluster have significant benefits in computation efficiency when compared to
other multi-polar approaches such as the GA method and DDSCAT.

\section{Results} \label{discussion_sect}

We present the results of the computation of the extinction of 
infrared, visible, and ultraviolet light by clusters of graphitic 
spheres of radii 10 nm and 50 nm, containing as few as 1 sphere to as
many as 343 spheres. Specifically, our results concern the wavelength
region from 0.1 to 100 $\mu$m. For convenience, we shall, from now 
onwards, call a cluster small if it has less than 10 particles, medium-size
if it has between 25 and 50 particles, and large if it has more than 100
particles. With this convention, we shall first discuss the convergence of 
the GA computations, followed by a comparison of the three computational
methods (GA, MarCoDES, and DDSCAT) when applied to the calculation of the
extinction of a single polycrystalline graphitic sphere. Then we shall 
present in succession our results for small, medium-size, and large clusters,
and end the section with an overall assessment of the computational methods
in light of the results presented.

\subsection{Convergence of GA method}

As we stated in Sect.\,\ref{ga}, with the GA method  we can compute
the extinction of a cluster in the $2^{L}$-polar approximation. We
find that 
the smallest L needed for the convergence of the extinction of our clusters
will be different for different regions of the optical spectrum. For 
small clusters, our full converged results show that in the UV-visible
it is sufficient
to use L = 7 for open clusters, and L = 9 
for compact ones, to compute the extinction with an accuracy of 1\%; 
whereas for longer wavelengths, at least L = 11 is required to ensure 
the same accuracy; this is so because in this region
graphite has a metallic-like behaviour  
(high $|m|$), and, thus, higher order multipoles have to be included 
in the computation of the extinction to get the same accuracy
as in the UV-visible range. We note that in the UV-visible range,
by accepting an accuracy 
of 5\% in the computation of the extinction, we can use L = 5 for 
open clusters and L = 7 for compact ones; we expect 
this to hold for clusters of up to a few tens of particles.  

In column 2 of Table\,\ref{table_ext}, we give the cut-off
polar order L used in the calculation of the extinction of all
our clusters. Full convergence was achieved only for small clusters;
for the others, L indicates the maximum polar order obtained with our 
computer resources. In general, L = 11 was used for small clusters,
L from 6 to 7 for medium-size clusters, and L from 2 to 3 for large
clusters.

\begin{figure}
   \centering
   \includegraphics[angle=90,width=9cm,clip]{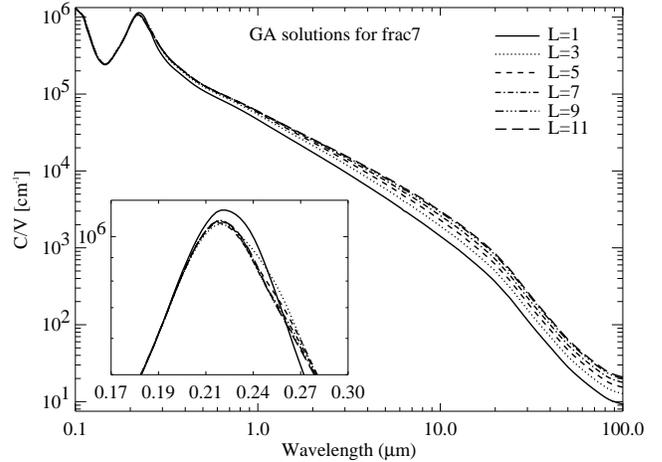}
      \caption{Solution of the GA 
   method for the fractal cluster structure containing 7 polycrystalline
   graphitic particles. 
   The solution is shown for different polar orders L. At L = 11 
    the solution was fully converged over the whole wavelength interval.
    The particle radius was 10~nm.}
         \label{GA_excat}
   \end{figure}

We show in Fig.\,\ref{GA_excat} the convergence of the extinction
of the frac7 cluster in the wavelength range $0.03-100~\mu$m.
The inset shows the 2175~{\AA} peak;
around the peak maximum L = 3 already gives results accurate within 1\%.

%
%                                                One column figure
%----------------------------------------------------------- 
%
%______________________________________________________________

\subsection{Single sphere}

To set up a comparison baseline, we compute the extinction of af single 
sphere of radius 10 nm using the two DDA codes (DDSCAT and MarCoDES) 
and the GA solution. Since the GA theory extends that of Mie (Sect.\ 4.1),
the GA and Mie solution coincide for a single sphere; this solution will 
be taken as a reference for the comparison.

\begin{figure}
\centering
\includegraphics[angle=90,width=9cm,clip]{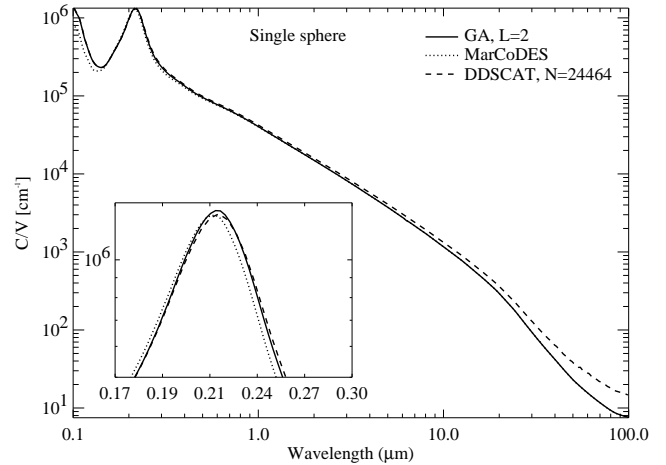}
\caption{A comparison between of DDSCAT, MarCoDES and the exact GA solution for 
a polycrystalline graphitic sphere with a 10 nm radius. 
The GA solution was fully converged at L = 2. 
The DDSCAT calculation was done with 24\,464 dipoles.}
\label{mie}
\end{figure}

It can be seen from Fig. \ref{mie} that DDSCAT 
is almost indistinguishable from the Mie solution for $\lambda \leq 1.0~\mu$m
while MarCoDES is so for
$\lambda \geq 0.6~\mu$m. The less good fit by DDSCAT for $\lambda > 1.0~\mu$m
is most likely a consequence of the refractive index of graphite having a
modulus larger than 2 for these wavelengths, and of the 24 464 dipoles being
inadequate to account for this.

%
%                                                One column figure
%-----------------------------------------------------------
%
%______________________________________________________________

\subsection{Small ($N < 10$) clusters}

\begin{figure}
   \centering
  \includegraphics[angle=90,width=9cm,clip]{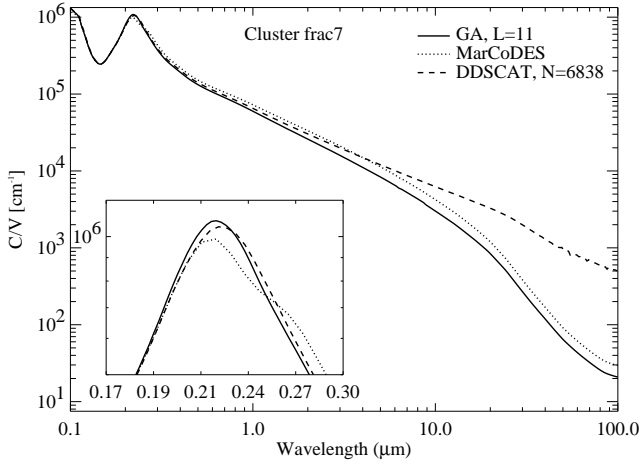}
      \caption{A comparison of the solutions from GA, MarCoDES and 
      DDSCAT of the extinction cross section of the fractal cluster,
      shown in Fig.\,\ref{frac}, consisting of 7 polycrystalline 
      graphitic particles.  The GA solution was fully converged at L = 11. 
     In the DDSCAT calculation 6\,838 dipoles were used. 
     The particle radius was 10 nm.}
         \label{frac7}
   \end{figure}

\begin{figure}
   \centering
  \includegraphics[angle=90,width=9cm,clip]{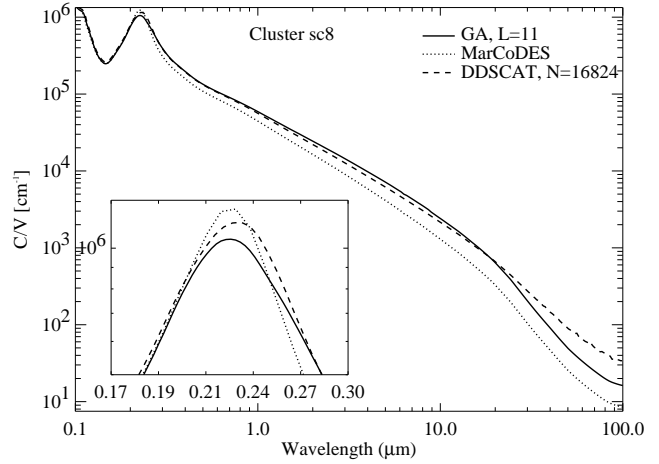}
      \caption{A comparison of the solutions from the
       GA, MarCoDES and DDSCAT calculations for the extinction
     cross section of the simple cubic cluster containing 8 polycrystalline graphitic
       particles.  The GA solution was fully converged at L = 11. 
     In the DDSCAT calculation 16\,824 dipoles were used. The particle
     radius was 10 nm.}
         \label{sc8}
   \end{figure}

As before, we take the fully converged GA solutions as a reference for comparison.
Fig.\,\ref{frac7} shows the extinction of the frac7 cluster as computed with
the three methods, while Fig.\,\ref{sc8} shows similar information for the sc8
cluster. 
For the frac7 cluster (Fig.\,\ref{frac7}) MarCoDES underestimates the extinction
for $0.2~\mu$m $< \lambda < 0.25~\mu$m and systematically 
overestimates it
for $\lambda > 0.26~\mu$m. DDSCAT\footnote{Calculated with
a $36 \times 36 \times 36$ dipole grid, which provides 6838 dipoles 
($\sim$ 977 dipoles per particle).} slightly underestimates the extinction for
$\lambda < 0.23~\mu$m and overestimates it for 
$\lambda > 0.23~\mu$m. In the range $0.26~\mu$m $< \lambda < 
4~\mu$m the excess extinction from DDSCAT is smaller than for the 
MarCoDES calculations while for $\lambda > 4~\mu$m DDSCAT gives a
systematic relative increase in the excess extinction with wavelength which is much
larger than the almost constant factor, $\sim 1.5$, seen for MarCoDES.  

%                                                One column figure
%----------------------------------------------------------- 

For the sc8 cluster (Fig.\,\ref{sc8}) the trend is different. 
DDSCAT\footnote{Calculated with a $32 \times 32 \times 32$ dipole 
grid, which provides 16\,824 dipoles ($\sim$ 2103 dipoles per particle).} 
overestimates the extinction for $\lambda < 0.3~\mu$m, underestimates it 
for $0.3~\mu$m $< \lambda < 11~\mu$m and overestimates it again for
$\lambda > 11~\mu$m.  MarCoDES also overestimates the extinction 
for $\lambda < 0.26~\mu$m but then systematically underestimates 
it to a much larger extent than DDSCAT for $\lambda > 0.26~\mu$m. 

%                                                One column figure
%----------------------------------------------------------- 
\begin{figure}
   \centering
  \includegraphics[angle=90,width=9cm,clip]{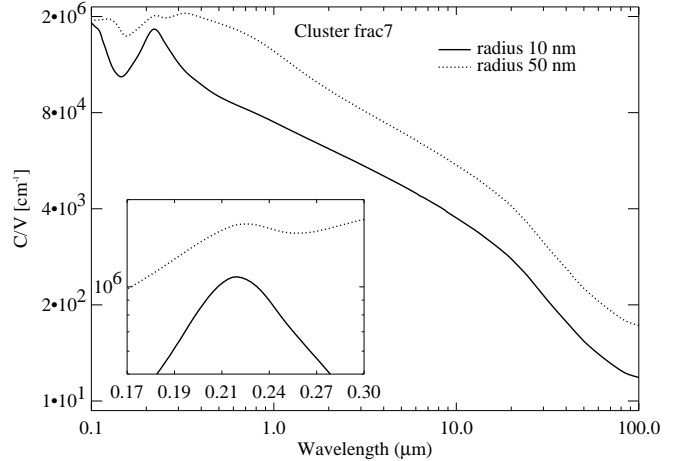}
      \caption{The extinction cross section obtained with GA for
      frac7 with particle radius of 10 and 50~nm. The calculations
       were fully converged at L = 11.}
         \label{frac7_50}
   \end{figure}

We have also investigated the effect of particle-size on a cluster's 
extinction using the fully converged results of the GA solution. For this, 
in addition to the extinction of the frac7 cluster of spheres of 
radii 10 nm, we have also computed the extinction of a frac7 cluster of 
spheres of radii 50 nm; which was fully 
converged at L = 11 with a precision of 1.2\%. Fig.\,\ref{frac7_50} 
shows the results; what 
strikes the eye immediately is the high increase in the extinction over 
almost the entire wavelength range investigated, upon increasing the radii 
of the spheres. Still, for wavelengths $\lambda > 1.1~\mu$m the shape of the 
extinction of both clusters look roughly the same. More important, 
however, are the differences observed in the wavelength range 
$\lambda < 1.1~\mu$, where the extinction of the cluster with the larger 
spheres exhibits peaks at $0.11~\mu$m and $0.4~\mu$m in addition to the 
2175~{\AA} peak, which is the only peak displayed by the extinction of 
the cluster with the smaller spheres. All these observations agree with 
the finding of Kimura (2001) that the light scattering of fractal 
clusters depends strongly on the size parameter of the monomer.  

%                                                One column figure
%----------------------------------------------------------- 

\subsection{Medium-size ($25 < N < 50$) clusters}

%                                                One column figure
%----------------------------------------------------------- 
   \begin{figure}
   \centering
   \includegraphics[angle=90,width=9cm,clip]{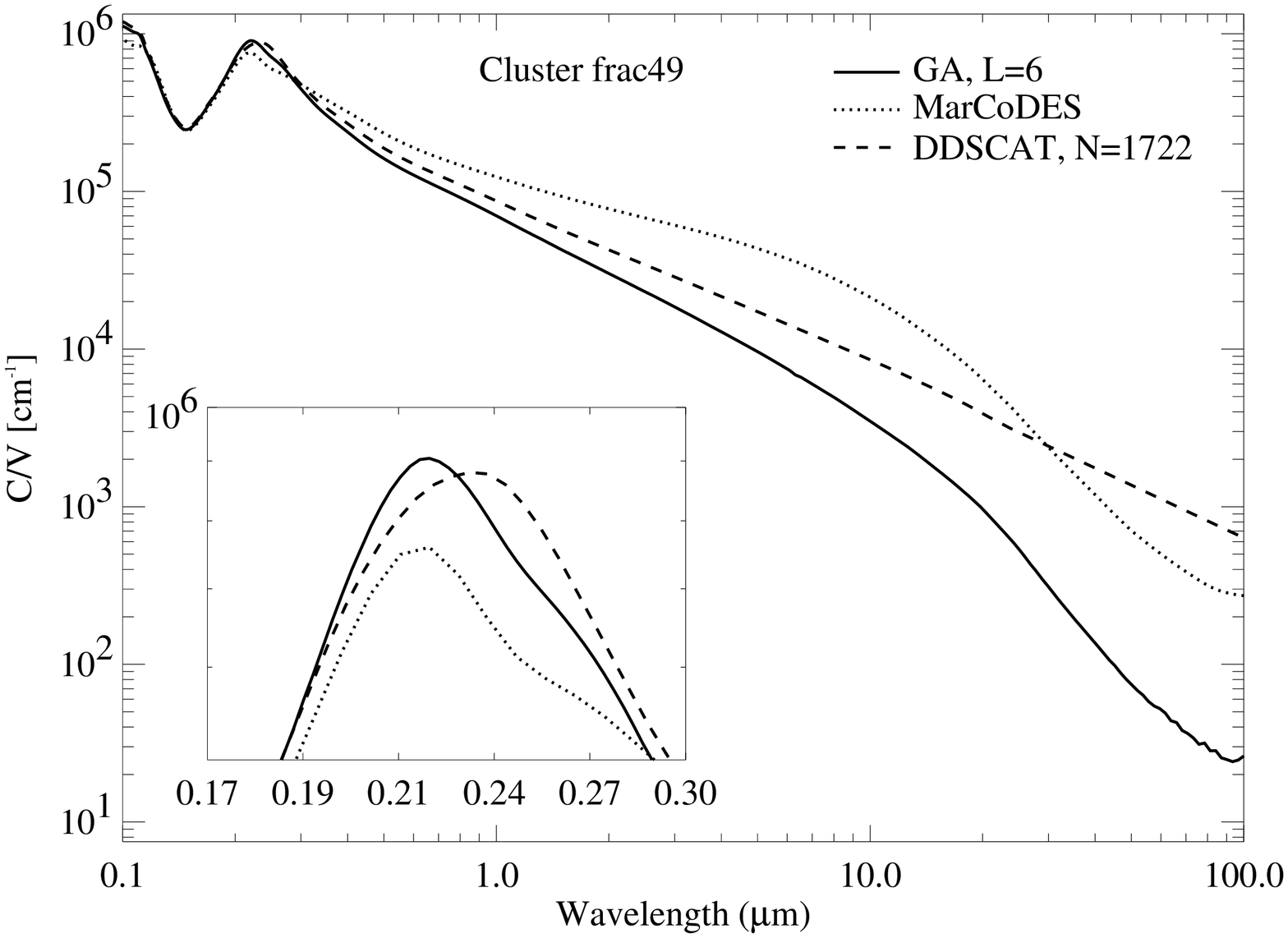}
      \caption{A comparison of the solutions from GA, MarCoDES and 
      DDSCAT for the extinction cross section of the frac49 cluster,
      shown in Fig.\,\ref{frac}. 
      The GA solution was truncated at L = 6. 
     In the DDSCAT calculation 1\,722 dipoles were used. 
     The particle radius was 10 nm.}
         \label{frac49}
   \end{figure}
%

%                                                One column figure
%----------------------------------------------------------- 
   \begin{figure}
   \centering
   \includegraphics[angle=90,width=9cm,clip]{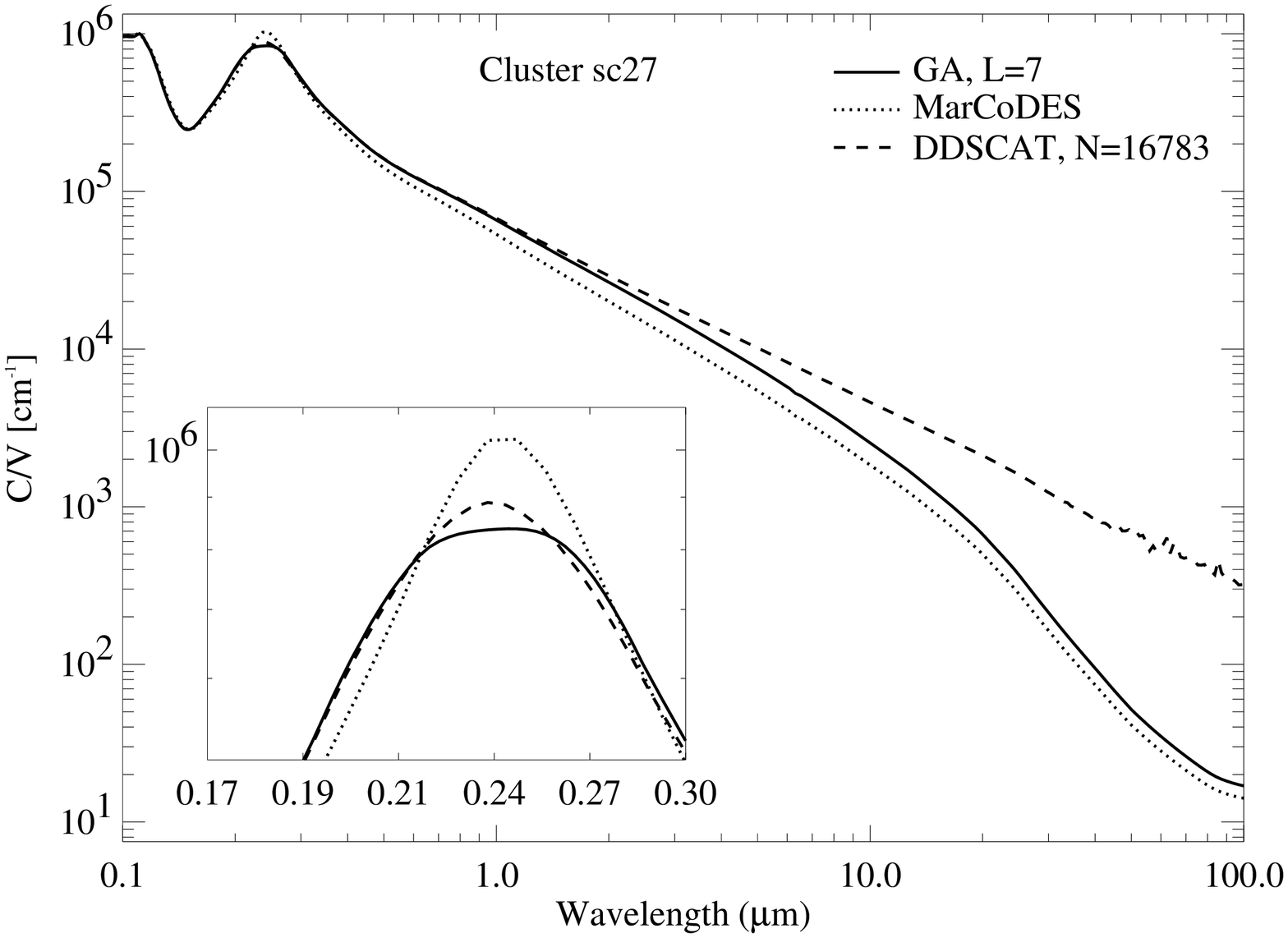}
      \caption{A comparison of the solutions from the
       GA, MarCoDES and DDSCAT calculations for the extinction
     cross section of the sc27 cluster.  
       The GA solution was truncated at L = 7. 
     In the DDSCAT calculation 16\,783 dipoles were used. The particle
     radius was 10 nm.}
         \label{sc27}
   \end{figure}
%

%                                                One column figure
%----------------------------------------------------------- 
   \begin{figure}
   \centering
   \includegraphics[angle=90,width=9cm,clip]{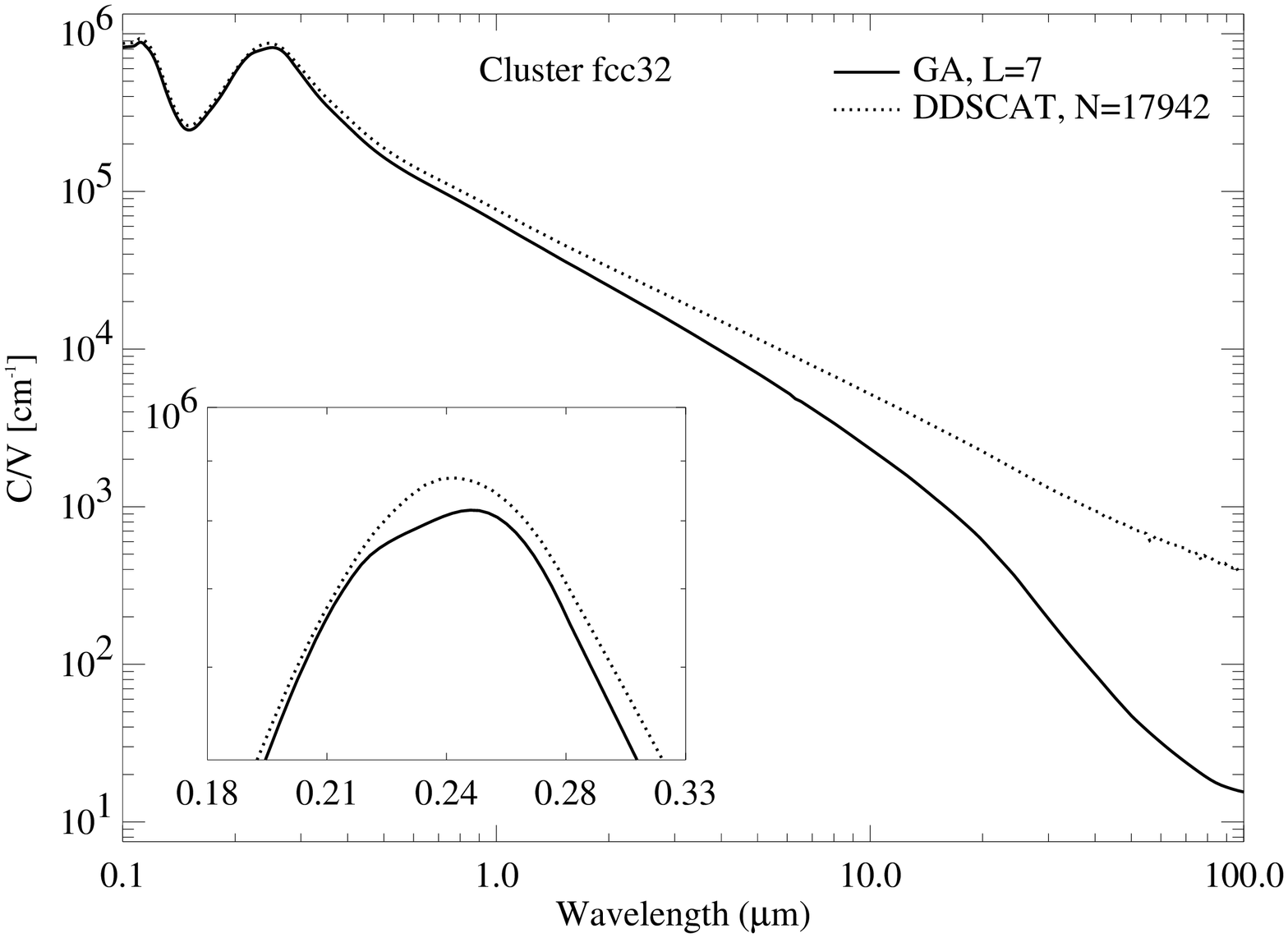}
      \caption{A comparison of the solutions from the
       GA and DDSCAT calculations for the extinction
     cross section of the fcc32 cluster.
      The GA solution was truncated at L = 7. 
     In the DDSCAT calculation 17\,942 dipoles were used. The particle
     radius was 10 nm.}
         \label{fcc32}
   \end{figure}

For the frac49 cluster, shown in Fig.\,\ref{frac}, the
extinction cross section obtained by the three methods 
is displayed in Fig.\,\ref{frac49}. 
For this structure neither DDSCAT\footnote{Calculated with a $36 \times 
36 \times 36$ dipole grid, which provides 1722 dipoles 
($\sim$ 35 dipoles per particle).} 
nor MarCoDES gets close to the solution suggested
from the GA calculations with L = 6. The GA
solution is not fully converged at long wavelengths 
due to computational limitations;
however, in the UV-visible region convergence is ensured at least
with an accuracy of 5\%.
Both MarCoDES and DDSCAT greatly
overestimates the extinction cross section for this cluster at longer wavelengths. 

As a comparison to the frac49 cluster we have carried out
calculations for two symmetric clusters, sc27 and fcc32, and an asymmetric
one, fcc49.

For the sc27 cluster (Fig.\,\ref{sc27}) the GA solution
was truncated at L = 7. MarCoDES underestimates the extinction 
for $0.17~\mu$m $< \lambda < 0.22~\mu$m and
$\lambda > 0.29~\mu$m. DDSCAT\footnote{Calculated with a $32 \times 32 \times 32$ 
dipole grid, this provides 16\,783 dipoles ($\sim$ 622 dipoles per particle).} 
also slightly underestimates the extinction for  $0.17~\mu$m $< \lambda < 0.22~\mu$m 
and $0.26~\mu$m $< \lambda < 0.4~\mu$m, but systematically overestimates it
for $ \lambda > 1~\mu$m.

We found that the version of MarCoDES tested here can not calculate a 
face-center cubic structure of touching particles because 
in this case the lattice cells representing neighbouring particles will touch
only at the corners, giving as a consequence a spectrum corresponding to
non-touching particles.
The extinction of fcc32 and fcc49 clusters were therefore only calculated 
with the GA theory and DDSCAT. The results for fcc32 are  displayed in 
Fig.\,\ref{fcc32}; the results for the fcc49 cluster look similar.
The GA calculations were truncated at L = 7 for fcc32 and
at L = 6 for fcc49.  For the DDSCAT calculations 
a $32 \times 32 \times 32$ dipole grid was used; for the fcc32 cluster
this provides 17\,942 dipoles ($\sim$ 561 dipoles per particle). For 
fcc49 a $48 \times 32 \times 40$ dipole grid was used; this
provides 29\,849 dipoles ($\sim$ 609 dipoles per particle).  
As can be seen in Fig\,\ref{fcc32}, 
DDSCAT systematically  overestimates the extinction especially at long 
wavelengths. The overestimation
of the extinction cross section as calculated by DDSCAT is a general tendency
for all the clusters we have studied. Also DDSCAT does not reproduce the absorption
hump expected for graphite around $10-15~\mu$m, which is
reproduced in the GA and MarCoDES calculations and also in Lorenz-Mie calculations
by Draine \& Li (\cite{draine+li}).

\subsection{Large ($N > 100$) clusters}

The GA solution could only be calculated to L = 3 for the fcc108 cluster
and to L = 2 for the frac343 cluster due to computational limitations.  
The extinction of the fcc108 and frac343 
clusters looks very similar to that of the fcc32 
and frac49 clusters shown in Figs.\,\ref{fcc32} and \ref{frac49}, respectively.
For the frac343 cluster the whole spectrum is shifted to longer
wavelengths by about 0.002~$\mu$m and 
for the fcc108 cluster the whole spectrum is shifted
0.01~$\mu$m to longer wavelengths compared to the frac49 and fcc32 cluster,
respectively. It was not possible to calculate these large clusters with DDSCAT.

\subsection{Discussion}

%                                                One column figure
%----------------------------------------------------------- 
    \begin{figure}
    \centering
   \includegraphics[angle=90,width=9cm,clip]{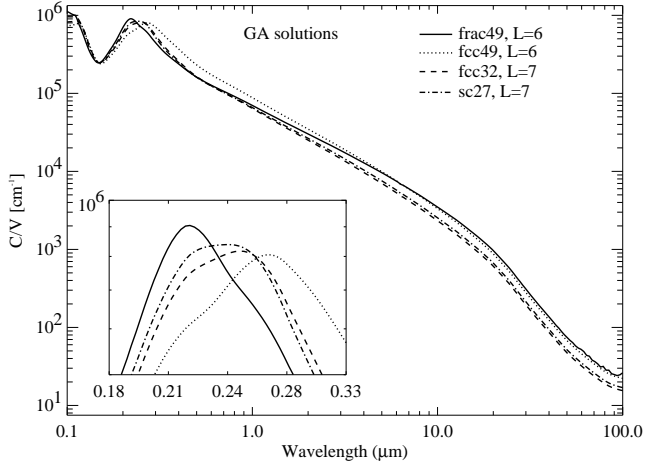}
      \caption{A comparison of the extinction cross section obtained with
     GA theory for the clusters frac49, fcc49, fcc32 and sc27. The 49 
      particle clusters were truncated at L = 6, while the two other
      clusters were truncated at L = 7. 
     The particle radius was 10 nm.}
         \label{GA_solutions}
   \end{figure}

It is seen that both DDA codes overestimate the extinction of the
fractal structures at long
wavelengths. A likely 
explanation is that the DDA model considers only 
point dipole interactions, whereas 
the GA solution matches the boundary conditions on the surface of
the spherical particle and solves the problem of
scattered and internal fields at the same time. Also, the large surface area
of fractal clusters contributes to the complexity of the scattered field, which
the GA method handles well because it  
takes into account the surface topology through the boundary conditions; 
the DDA codes, on the other hand, perform best when the surface to volume
ratio is low. 
This interpretation is also supported by the calculations for the 
small compact sc8 cluster, see Fig.\,\ref{sc8}, where the
DDA approximations perform reasonably well. For larger compact
clusters, however, DDSCAT does not give good results at long
wavelengths. This is due to computational limitations; the number
of dipoles per particle that we could handle becomes lower as
the cluster size increases. It is interesting to note that
MarCoDES gives better results for compact than for fractal 
cluters of intermediate size.
According to Markel et al. (\cite{markel00}), 
the performance of MarCoDES can be improved by altering
the intersection parameter which determines if the particles touch or
overlap, but we have not investigated that here. Nevertheless, 
the fact that MarCoDES in some
cases overestimates the extinction, as compared to GA theory, and in 
other cases underestimates it, suggests that the determination of
the rather arbitrary
optimal intersection parameter is a very complex problem indeed.

We next compare the extinction of fractal and compact clusters of
intermediate size.  In Fig.\,\ref{GA_solutions}, the 
GA calculation for frac49 is compared to that for fcc49 and fcc32. 
It is seen that the extinction of fractal and compact clusters are of the 
same order of magnitude. This is in contradiction to the DDA
calculations which give a much higher extinction for fractal clusters
(Figs.\,\ref{frac7} - \ref{GA_solutions}).
We, therefore, recommend that great care should be taken when using these
DDA codes at longer wavelengths for fractal clusters with low fractal dimension
consisting of materials with high refractive index ($m$).

Concerning the effect of particle size on the extinction cross
section, in classical electromagnetic theory the extinction is independent
of cluster size in the long wavelength limit, i.e. when the size is much
smaller than the wavelength. As seen in Fig.\,\ref{frac7_50},
clusters consisting of particles of radii 50 nm are
clearly outside of this limit. Decreasing the
particle radius below 10 nm, however, will give, in a classical theory, 
at most minor
effects on the extinction of single particles and small clusters.
Quantum mechanical effects, on the other hand, may influence the width and
the position of the absorption peak for such small particles. A detailed
study of these particle size effects on the extinction cross section is
outside the scope of the present work.

\section{The 2175~{\AA} extinction peak} \label{fuv_sect}

The main observational constraints concerning the 2175~{\AA} peak 
are according to Fitzpatrick \& Massa (\cite{fitzpatrick+massa86})
and Mathis (\cite{mathis}):
\begin{enumerate}
\item The peak position is remarkably constant, $\nu_{\rm max} = 4.6 \pm 0.04~\mu$m$^{-1}$, where $\nu = 1 / \lambda$.
\item The peak width has a much larger range of variations of $1.0 \pm 0.25~\mu$m$^{-1}$ (i.e. $\le 25\%$).
\item The variation in peak position and the width ($\gamma$) 
are uncorrelated, except for
the widest bumps 
($\gamma \ge 1.2~\mu$m$^{-1}$) where a systematic shift to larger
wavenumber is observed 
(Cardelli \& Savage \cite{cardelli+savage}; Cardelli \& Clayton 
\cite{cardelli+clayton}). These later observations all have lines of sight 
passing through dark, dense regions.
\end{enumerate}

%                                                One column figure
%----------------------------------------------------------- 
   \begin{figure}
   \centering
   \includegraphics[angle=90,width=9cm,clip]{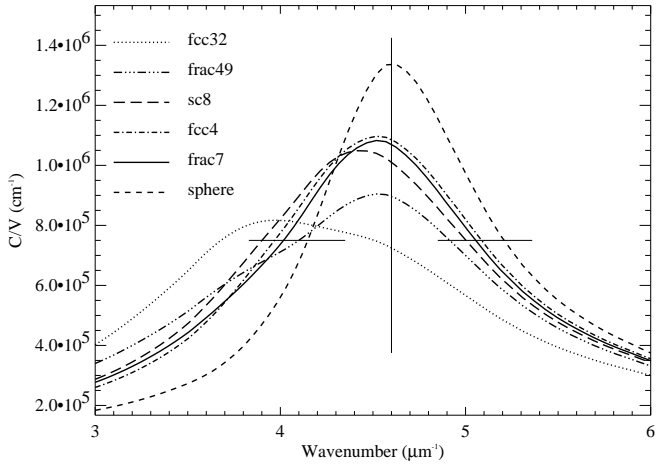}
      \caption{Extinction cross section for
    polycrystalline graphitic clusters of different sizes calculated with the
      GA method. A difference in peak position can be seen 
      between the fractal clusters and compact clusters.
    The lines in the plot indicate the
     peak position and width variation of the interstellar extinction curve.
    All clusters consist of particles with a 10 nm radius. Also shown is the 
    GA solution for a sphere, radius 10~nm, which is 
      equivalent to the Lorenz-Mie solution.}
         \label{ext}
   \end{figure}

\begin{table}
\begin{center}
\caption{Peak position and FWHM of different clusters, as calculated
with the GA method. The value of L indicates at which polar
order the GA calculations were truncated.}
\begin{tabular}{|c|c|c|c|} \hline
cluster name & L & peak & FWHM \\ 
  &  & [$\mu$m$^{-1}$] &  [$\mu$m$^{-1}$] \\ \hline
 frac7 & 11 & 4.52 & 1.17  \\
 frac49 & 6 & 4.52 & 1.33 \\
 frac343 & 2 & 4.42 & 1.27 \\
 sphere & 3 & 4.62 & 0.96\\
 fcc4 & 11 & 4.43 & 1.15 \\
 fcc32 & 7 & 3.98 & 1.43 \\
 fcc 49 & 6 & 3.71 & 1.52 \\
 fcc108 & 3 & 3.98 & 2.01 \\ 
 sc8 & 11 & 4.41 & 1.27  \\ 
 sc27 & 7 & 4.12 & 1.62 \\ \hline
\end{tabular}
\label{table_ext}
\end{center}
\end{table}
     
Table \ref{table_ext} lists the peak position and width for the 
different cluster structures, as computed with the GA method.
The polar order at which the GA calculation was truncated is
also indicated. The width was determined as the FWHM. For 
the asymmetric compact cluster, we have investigated
the importance of the orientation of the cluster and it was found
that the peak position was shifted less than 0.1 $\mu$m$^{-1}$ while
the width of the feature was not affected. 

It was
shown in a study  by Rouleau et al. (\cite{rouleau}), that small compact
clusters qualitatively satisfy the observational constraints on the
2175~{\AA} peak, except that the peak position
falls at the wrong wavelength. This result is not confirmed from
our findings, since the compact clusters we have studied do not have
a stable peak position. 
The clusters considered by Rouleau et al. (\cite{rouleau}) all
had peak positions at higher wavenumber than the 2175~{\AA} peak 
while the clusters we have considered all have
peak position at lower wavenumbers.
Rouleau et al. (\cite{rouleau}) also used optical constants from Draine \&
Lee (\cite{draine+lee}), but they dealt with the anisotropy of the
graphitic particles with the usual ``$1/3-2/3$'' approximation, see 
Sect. \ref{graphite_sect}, which
assumes the particles to be mono-crystalline. As
argued in Sect.\,\ref{graphite_sect} assuming the particles to be
polycrystalline seems much more reasonable. 
This implies that variations in crystallinity is indeed an important factor
when discussing the 2175~{\AA} peak as also suggested by Draine \& Malhotra
(\cite{draine+malhotra}).

The fractal clusters considered here do not enhance 
the long wavelength wing and do not introduce additional structure as seen by 
Rouleau et al. (\cite{rouleau}) for elongated and "fluffy" structures. 
The frac7 and frac49 clusters do come close (peak position is
off by $0.04-0.08~\mu$m$^{-1}$) to the observational constraints.
This indicates that small ($N \sim 5 - 100$) 
fractal clusters ought to be investigated in more detail
to determine if fractal clusters with low fractal
dimension have a stable peak position around $4.6~\mu$m$^{-1}$ 
and produce a variable width depending on the number of particles in the cluster.

\section{Conclusions} \label{conclusion_sect}

We computed the extinction of clusters of polycrystalline graphitic
particles in the wavelength range $0.1$ to $100~\mu$m.  Computations by the
rigorous multipolar theory of G\'erardy \& Ausloos (\cite{GA}; GA) were
compared to two formulations of the discrete dipole approximation (DDA). 

We have compared the extinction of open fractal clusters and compact clusters.
The fractal and compact clusters display an extinction at long wavelengths,
of the same order of magnitude as computed with the GA method. It seems 
that the DDA approximations grossly overestimate the long-wavelength 
extinction of small fractal structures. If this also holds true for 
larger fractal clusters is not possible for us to say due to computational 
limitations. The DDA codes should, however, be used with caution for this 
type of problem.

The GA computations were compared to the observed interstellar extinction
curve. We found that results for small and medium-size (less than 50 particles)
fractal clusters are in fair agreement with observational constraints, while those of
compact sc and fcc clusters are not. 

Convergence of the GA computations for small (less than 10 particles) clusters
was found by including 11 multipoles. Good approximative results were
obtained for medium-size (between 25 and 50 particles) 
clusters. In most cases we
found a substantial differences between the GA theory and the DDA
approximations of Draine \& Flatau (\cite{draine+flatau}; DDSCAT)
and Markel (\cite{markel}; MarCoDES).

As predicted by Draine \& Goodman (\cite{draine+goodman}) the absorption
is significantly overestimated by DDSCAT at the wavelengths where the 
modulus of the refractive index, $|m|$, is large, 
see Figs.\,\ref{frac49}, \ref{sc27} and \ref{fcc32}. 
For $|m|<2$, DDSCAT performs well and
better than the MarCoDES. MarCoDES, however, is computationally much faster than 
DDSCAT and the GA calculations.  With DDSCAT the accuracy is
directly dependent on the number of dipoles used in the approximation. Hence,
by using more dipoles than we have used here, a better accuracy 
could be obtained, but then the computational cost would increase.
Which DDA code is best to use will
therefore depend on the type of problem one wants to address and the accuracy 
needed.

\begin{acknowledgements}
      We would like to thank B.T.\,Draine and V.A.\,Markel for making their
      DDA codes available as shareware. VP would like to thank 
       V.A.\,Markel and ACA would like to thank Bj\"orn Davidsson 
     for fruitful discussions.
      ACA acknowledges support from the Carlsberg Foundation and from NorFA.
      JS and GN acknowledges support from the Swedish Natural Science Research Council (NFR). 
      VP acknowledges support from the Wenner-Gren Foundation.
\end{acknowledgements}


\begin{thebibliography}{}


   \bibitem[1988]{avellaneda} Avellaneda M., Cherkaev A.M., Lurie K.A. \& Milton G.W., 1988,
       J.\ Appl.\ Phys.\ 63, 4989

   \bibitem[1990]{bazell} Bazell D. \& Dwek E., 1990, ApJ 360, 262

   \bibitem[1986]{berry} Berry M.V. \& Percival I.C., 1986, Opt.\ Acta 33, 577

   \bibitem[1983]{bohren+huffman} Bohren C.F. \& Huffman D.R., 
       1983, Absorption and Scattering
       of Light by Small Particles (John Wiley \& Sons, New York)

   \bibitem[1991]{borghesi+guizzetti} Borghesi A. \& Guizzetti G., 1991, in Handbook of optical
      constants of solids II, ed.\ E.D. Palik (Academic Press, New York), 449

    \bibitem[1988]{botet} Botet R. \& Jullien R., 1988, Ann.\ Phys.\ (France) 13, 153

   \bibitem[1988]{cardelli+savage} Cardelli J.A. \& Savage B.D., 1988, ApJ 345, 245

   \bibitem[1991]{cardelli+clayton} Cardelli J.A. \& Clayton G.C., 1991, AJ 101, 1021

    \bibitem[2000]{Ciric} Ciric I.R. \& Cooray F.R., 2000, in
    Light Scattering by Nonspherical Particles: 
      Theory, Measurements and Applications,
   eds. M.I. Mishchenko, J.W. Hovenier, L.D. Travis
    (Academic Press, New York), 89

   \bibitem[1994]{daulton} Daulton T.L., Eisenhour D.D., Lewis R.S \& Bernatowicz T.J., 
         1994, Lunar Planet.\ Sci.\ 25, 313

   \bibitem[1909]{debye} Debye P., 1909, Ann.\ Phys.\ 30, 57

   \bibitem[1988]{draine88} Draine B.T., 1988, ApJ 333, 848

   \bibitem[2000]{draine00} Draine B.T., 2000, 
     in Light Scattering by Nonspherical Particles: Theory, 
    Measurements, and Applications, ed.\ M.I. Mishchenko, 
    J.W. Hovenier \& L.D. Travis (Academic Press, New York), 131

   \bibitem[1994]{draine+flatau} Draine B.T. \& Flatau P.J., 1994, J.\ Opt.\ Soc.\ Am.\ A 11, 1491

   \bibitem[2000]{draine+flatau00} Draine B.T. \& Flatau P.J., 2000, User guide 
     for the Discrete Dipole Approximation Code
   DDSCAT (Version 5a10), http://xxx.lanl.gov/abs/astro-ph/0008151v3

   \bibitem[1993]{draine+goodman} Draine B.T. \& Goodman J.J., 1993, ApJ 405, 685
 
   \bibitem[2001]{draine+li} Draine B.T. \& Li A., 2001, ApJ 554, 778

   \bibitem[1984]{draine+lee} Draine B.T. \& Lee H.M., 1984, ApJ 285, 89

   \bibitem[1993]{draine+malhotra} Draine B.T. \& Malhotra S.K., 1993, ApJ 325, 864
    
   \bibitem[1986]{fitzpatrick+massa86} Fitzpatrick E.L. \& Massa D., 1986, ApJ 307, 286

   \bibitem[1988]{fitzpatrick+massa} Fitzpatrick E.L. \& Massa D., 1988, ApJ 328, 734

   \bibitem[2000]{Fuller} Fuller, K.A. \& Mackowski, D.W, 2000, in
    Light Scattering by Nonspherical Particles: 
      Theory, Measurements and Applications,
   eds. M.I. Mishchenko, J.W. Hovenier \& L.D. Travis (Academic Press, New York), 225
      
   \bibitem[1980]{GA4} G\'erardy J.M. \& Ausloos M., 1980, Phys.\ Rev.\ B 22, 4950

   \bibitem[1982]{GA} G\'erardy J.M. \& Ausloos M., 1982, Phys.\ Rev.\ B 25, 4204

   \bibitem[1983]{GA2} G\'erardy J.M. \& Ausloos M., 1983, Phys.\ Rev.\ B 27, 6446

   \bibitem[1984]{GA3} G\'erardy J.M. \& Ausloos M., 1984, Phys.\ Rev.\ B 30, 2167

   \bibitem[1972]{gilra} Gilra D.P., 1972, in The Scientific Results of 
      OAO-2, ed. A.D. Code (NASA, SP310), 295

   \bibitem[1991]{goodman} Goodman J.J., Draine B.T. \& Flatau P.J., 1991, Opt.\ Lett.\ 16, 1198

   \bibitem[1990]{hage+greenberg} Hage J.I. \& Greenberg J.M., 1990, ApJ 361, 251

   \bibitem[1968]{harrington} Harrington R.F., 1968, 
       Field Computation by Moment Methods, (Macmillan, New York)

   \bibitem[1981]{hecht} Hecht J., 1981, ApJ 246, 794
 
   \bibitem[2001]{kimura} Kimura H., 2001, Journal of Quantitative Spectroscopy
    \& Radiative Transfer 70, 581

   \bibitem[1986]{kittel} Kittel C., 1986, Introduction to Solid State Physics, 
        (Wiley, New York), 11

   \bibitem[1999]{lebedev1} Lebedev A.N. \& Stenzel O., 1999, Eur.\ Phys.\ J.\ D 7, 83
  
   \bibitem[1999]{lebedev2} Lebedev A.N., Gartz M., Kreibig U. \& Stenzel O., 
     1999, Eur.\ Phys.\ J.\ D 6, 365

   \bibitem[1890]{lorenz} Lorenz L., 1890, Vidensk. Selsk. Skr. T. VI(6),
       (Bianco Lunos Kgl. Hof-Bogtrykkeri, Copenhagen), 1

   \bibitem[1983]{mandelbrot} Mandelbrot B.B., 1983, 
       The Fractal Geometry of Nature (Freeman, San Francisco)

   \bibitem[1998]{markel} Markel V.A., 1998, User guide for 
     MarCoDES - Markel's Coupled Dipole Equation Solvers, 
     http://atol.ucsd.edu/~pflatau/scatlib/  

    \bibitem[2000]{markel00} Markel V.A., Shalaev V.M. \& George T.F., 2000,
     in Optics of Nanostructured Materials, eds. V.A. Markel \& T.F. George
     (Wiley, New York), 355

   \bibitem[1994]{mathis} Mathis J.S., 1994, ApJ 422, 176

   \bibitem[1988]{meakin} Meakin P., 1988, Ann.\ Rev.\ Phys.\ Chem.\ 39, 237

   \bibitem[1988]{meakin+jullien} Meakin P. \& Jullien R., 1988, J.\ Chem.\ Phys.\ 89, 246

   \bibitem[1908]{mie} Mie, G., 1908, Ann.\ Phys.\ Leipz.\ 25, 377

   \bibitem[2000a]{Mish1} Mishchenko M.I., Wiscombe W.J, Hovenier J.W. \& 
    Travis L.D., 2000a, in Light Scattering by Nonspherical Particles: 
      Theory, Measurements and Applications, 
       eds. M.I. Mishchenko, J.W. Hovenier \& L.D. Travis 
       (Academic Press, New York), 29

   \bibitem[2000b]{Mish2} Mishchenko M.I., Travis L.D. \& Macke A., 2000b, 
    in Light Scattering by Nonspherical Particles: 
      Theory, Measurements and Applications,
   eds. M.I. Mishchenko, J.W. Hovenier \& L.D. Travis
    (Academic Press: New York), 147
   
   \bibitem[1999]{phelps} Phelps A.W., 1999, Lunar Planet.\ Sci.\ 30, 1749
 
   \bibitem[1973]{purcell} Purcell E.M. \& Pennypacker C.R., 1973, ApJ 186, 705

   \bibitem[1997]{rouleau} Rouleau F., Henning Th. \& Stognienko R., 1997, A\&A 322, 633

   \bibitem[1975]{ruppin} Ruppin R., 1975, Phys.\ Rev.\ B 11, 2871

   \bibitem[1994]{sedlmayr} Sedlmayr E., 1994, in Molecules in the Stellar Environment,
     LNP 428, ed.\ U.G. J{\o}rgensen (Springer, Berlin), 163

  \bibitem[1995]{snow+witt} Snow T.P. \& Witt A.N., 1995, Sci.\ 270, 1455 

   \bibitem[1990]{sorell} Sorell W.H., 1990, MNRAS 243, 570

   \bibitem[1965]{stecher} Stecher T.P., 1965, ApJ 142, 1683

   \bibitem[1965]{stecher+donn} Stecher T.P. \& Donn B., 1965, ApJ 142, 1681 

   \bibitem[1985]{tsang}Tsang L, Kong J.A. \& Shin R.T., 1985, Theory of Microwave Remote Sensing 
    (Wiley, New York), chapter 3 

   \bibitem[1997]{xing+hanner} Xing, Z. \& Hanner M.S., 1997, A\&A 324, 805

  \bibitem[1999]{xu+gustafson} Xu Y-L. \& Gustafson B.{\AA}.S., 1999, ApJ 513, 894

   \bibitem[2001]{vaidya} Vaidya D.B., Gupta R., Dobbie J.S. \& Chylek P., 2001, A\&A 375, 584

   \bibitem[1983]{vicsek} Vicsek T., 1983, J.\ Phys.\ A 16, L647

   \bibitem[1990]{nicolai90} Voshchinnikov N.V., 1990, Sov.\ Astron.\ Lett.\ 16, 215

   \bibitem[2002]{nicolai02} Voshchinnikov N.V., 2002, Astroph.\ \& Space Sci.\ Rev.\ 12, 1 

   \bibitem[1971]{waterman}Waterman P.C., 1971, Phys.\ Rev.\ D 3, 825

   \bibitem[1999]{will+aannestad} Will L.M. \& Aannestad P.A., 1999, ApJ 526, 242
   
   \bibitem[1998]{wurm} Wurm G. \& Blum J., 1998, Icarus 132, 125


\end{thebibliography}
\end{document}